\documentclass[twocolumn]{aastex631}

\usepackage[mathscr]{euscript}
\usepackage{amsmath}

\begin{document}

\title{Little Red Dots are  Tidal Disruption Events in Runaway-Collapsing Clusters}

\correspondingauthor{Jillian Bellovary}
\email{jbellovary@amnh.org}

\author[0000-0001-7596-8372]{Jillian Bellovary}
\affiliation{Department of Physics, Queensborough Community College, 222-05 56th Ave, Bayside, NY 11364, USA}
\affiliation{Astrophysics Program, CUNY Graduate Center, 365 5th Ave, New York, NY 10016, USA}
\affiliation{Department of Astrophysics, American Museum of Natural History, New York, NY 10024, USA}

\begin{abstract}

I hypothesize a physical explanation for the ``Little Red Dots" (LRDs) discovered by the James Webb Space Telescope (JWST).  The first star formation in the universe occurs in dense clusters, some of which may undergo runaway collapse and form an intermediate mass black hole.  This process would appear as a very dense stellar system, with recurring tidal disruption events (TDEs) as stellar material is  accreted by the black hole.  Such a system would be compact, UV-emitting, and exhibit broad H$\alpha$ emission.  If runaway collapse is the primary mechanism for forming massive black hole seeds, this process could be fairly common and explain the large volume densities of LRDs.   In order to match the predicted number density of runaway collapse clusters, the tidal disruption rate must be on the order of  $10^{-4}$ per year.   A top-heavy stellar initial mass function may be required to match observations without exceeding the predicted $\Lambda$CDM mass function.  The TDE LRD hypothesis can be verified with followup JWST observations looking for TDE-like variability.

\end{abstract}


\section{Introduction} \label{sec:intro}

The objects known as Little Red Dots (LRDs) discovered by the James Webb Space Telescope (JWST) have been mystifying our community \citep{Greene24,Kocevski24,Kokorev24,Matthee24}.  They exhibit features common to both star-forming galaxies and active galactic nuclei (AGNs), but their emission cannot be modeled robustly by either (or a combination).  Their number density at high redshift is also higher than expected compared to similar galaxy populations.  If they are dominated by star formation, they are more compact than anything ever seen before \citep{Guia24}, and exceed the predicted stellar mass function allowed by $\Lambda$CDM \citep{MBK23,Baggen24,Akins24}.  If they are dominated by AGN emission, their existence implies an almost 100\% AGN fraction in high-redshift galaxies \citep{Greene24}.  Both of these possibilities are difficult to reconcile with any  cosmic evolution model.

LRDs have been discovered mostly at redshifts $5 < z < 8$, and often exhibit a ``V-shaped'' spectral energy distribution (SED).  The bottom of the ``V'' is at the Balmer break \citep{Setton24}, with rising luminosity both in the UV and red wavelengths.  They have broad H$\alpha$ emission lines but the broadness is not seen in H$\beta$ \citep{Brooks24,Maiolino24}.   They are X-ray undetected or weak \citep{Yue24} and are also undetected in the far-infrared \citep{Labbe23,Perez-Gonzalez24,Williams24}.  This lack of emission not only makes the SEDs very difficult to characterize and bolometric luminosities hard to determine, but there is an overall struggle to pinpoint the source of the emission at all.

One of the many reasons LRDs are fascinating is their potential to shed light on the formation and evolution of supermassive black holes (SMBHs).  The formation mechanism of SMBH seeds, often thought to be intermediate mass black holes (IMBHs) of some variety, is highly unconstrained, as is their subsequent growth into SMBHs.  Seeds may be ``light'' (i.e. the remnants of the first stars, e.g. \citet{Bromm03}), ``heavy'' (i.e. direct collapse black holes,  e.g. \citet{Begelman06}), or in-between (i.e. remnants of runaway cluster collapse, e.g. \citet{Devecchi09}), and are likely to undergo at least some phases of super-Eddington growth to reach supermassive size.  Observations of the earliest galaxies could untangle this quandary - or at least we thought so.

 In this Letter I explore the hypothesis that Little Red Dots can be explained by black hole seeds in-the-making via tidal disruptions in  runaway collapsing clusters. 
The resulting SED would somewhat resemble an AGN but with some key differences, many of which are consistent with LRD observations.

\section{Runaway Collapse}\label{sec:cc}

One prominent theory for the formation of black hole seeds is the runaway collapse of dense star clusters.  It is thought to create seeds of $\sim 1000 M_\odot$ and may or may not be redshift dependent, depending on whether metallicity is an important factor.  The phenomenon of runaway collapse involves a dynamical instability resulting in the collapse of the cluster core.    The ensuing runaway collisions  in these nuclear star clusters may result in the creation of a supermassive star \citep{PortegiesZwart99,Ebisuzaki01,PortegiesZwart04,Devecchi09} which then evolves into an intermediate mass black hole.    This phenomenon has been demonstrated in Monte Carlo  \citep{Gurkan04,Freitag06} and N-body \citep{Rantala25} simulations, and runaway collapse may be enhanced in cluster with low metallicities and higher masses \citep{DiCarlo21} and/or high binary fractions of massive stars \citep{Gonzalez21}.  Other works suggest IMBH formation may involve a pre-existing black hole which rapidly accretes surrounding objects in the cluster  \citep{Begelman78,Davies11}. Regardless of the precise order of operations, a dense cluster surrounding a massive black hole will be an efficient environment for tidal disruption events.  In fact, \citet{Alexander17} suggest that intermediate mass black holes should not exist at all because growth from TDEs is so efficient that the minimum MBH mass is $10^5 M_\odot$.

\begin{figure}
\includegraphics[width=0.45\textwidth]{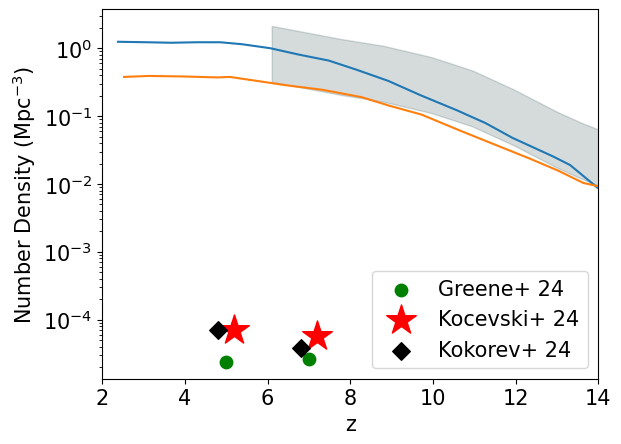}
\caption{ Comoving number density of runaway collapse black hole seeds vs redshift.  Predictions from numerical simulations vary based on whether a delayed cooling model is used with supernova feedback (blue line) or not (orange line) \citep{Habouzit17}.  Analytic predictions using variable seed masses are shown in the grey shaded area \citep{Devecchi12}.  Data points represent JWST LRD observations from \citet{Greene24} (green dots), \citet{Kocevski24} (red stars) and \citet{Kokorev24} (black diamonds).  The data are $\sim 4$ orders of magnitude below the predicted densities.
\label{fig:lrd_density}
}
\end{figure}

There have been very few calculations of the number density of such seeds with cosmic time, likely because the conditions for runaway collapse happen on a much smaller scale than that of e.g. cosmological simulations.  Estimates have been made by \citet{Devecchi12} (analytic) and \citet{Habouzit17} (numerical)  which I show in Figure  \ref{fig:lrd_density}.  The grey shaded area represents an adaptation of the prediction from \citet{Devecchi12} for a range of seed masses ($300 - 3000 M_\odot$).   The solid lines are from a cosmological hydrodynamic simulation using a subgrid model  with star-formation-based IMBH formation \citep{Habouzit17}.  The blue line shows a model with delayed cooling after supernova explosions, while the orange represents regular kinetic or thermal supernova feedback.  
  Their model includes a combination of a cluster-based model,  which is based on the formation of a supermassive star as described in the previous paragraph,  and seeds from Population III stars.    This model specifically aims to create seeds based on the formation of the first stars and  nuclear star clusters.  Both types of seeds are formed by finding dense pristine clumps of gas with an average mass of $\sim 1000$M$_\odot$.   Their results do not differentiate the two threads of seed formation, so the data in Figure \ref{fig:lrd_density} may be taken as an upper limit.   
Also included in this figure are observational results of LRDs from JWST: \citet{Greene24} (green dots), \citet{Kocevski24} (red stars) and \citet{Kokorev24} (black diamonds).  The observed points are about 4 orders of magnitude below the predicted density of black hole seeds formed through cluster collapse.  A cannonical TDE has a peak bolometric luminosity of $\sim 10^{44}$ erg s$^{-1}$ and remains observable for about 1 year \citep{Gezari21}.  Therefore, a TDE rate of $10^{-4}$ per year could explain the fraction of observed LRDs vs the occupation of black holes in galaxies.

\section{TDE Rates in LRDs} \label{sec:tde}

Tidal disruption events happen when stars enter the tidal radius of a black hole.  As a result, approximately half of the star's mass becomes unbound, and the remainder  is stretched into streams before being eventually devoured by the black hole.  These events likely result in super-Eddington accretion, and are UV bright but often X-ray weak.  Broad H$\alpha$ lines with widths of $3-13 \times 10^4$ km s$^{-1}$ are common \citep{Gezari21}.  Their brightness decays with a characteristic $t^{-5/3}$ powerlaw and lasts on the order of 100's of days.  I now quantify whether the TDE rate estimated in the previous section is physically reasonable for powering black hole growth.

\citet{Stone17} suggests that a combination  of tidal capture and TDEs can grow a black hole quickly and efficiently.  They give the following rate:

\begin{equation}
\dot N = n_\star \Sigma v_{rel}
\end{equation}

where  the number density of stars $ n_\star = \rho / M_\star $ with a mean stellar mass of $M_*$,   $v_{rel} $ is the relative velocity between the MBH and the star (which I take to equal the velocity dispersion $\sigma$), and $\Sigma$ is the tidal-capture cross section, given by 

\begin{equation}
\Sigma = \pi R_t^2 \left( 1 + \frac{2 G M_{tot} }{R_t v_{rel}^2} \right).
\end{equation}

Here $M_{tot}$ is the total mass ($M_{BH}+M_*$), and $R_t$ is the tidal radius given by $R_t = r_\star \left(M_{BH}/M_\star \right)^{1/3}$ and $r_\star$  is the solar radius multiplied by $M_*^{0.8}$.  Tidal capture is likely dominant at lower black hole masses (e.g. $M < 10^3 M_\odot$) and would contribute minimal luminosity.    A possible limitation of this model is that initial black hole growth may be delayed because tidally captured stars will inflate before being consumed, causing a slow build before runaway growth occurs.  Therefore, it is uncertain whether this method will be fast enough to jumpstart the growth of  LRD black holes by the time they are observed.  As the dominant process shifts to TDEs at higher masses, \citet{Stone17} argue that in the full loss cone approximation, the growth rate would remain similar and still scale as $M_{\rm BH} ^ {4/3}$.  However, it is also unclear at which point, if any, the growth of the black hole transitions from being dominated by tidal captures to tidal disruption events.    The evolution of a collapsing cluster is dynamic, and may include further gas inflow from the larger environment (and subsequent star formation), so it is likely that a full loss cone is an appropriate assumption.

In addition to the prior analytic estimate, I consider a rate developed using numerical simulations.  \citet{Rizzuto23} use the BIFROST code \citep{BIFROST} to study the growth of a black hole in a cluster  direct N-body methods including Post-Newtonian dynamics and a detailed TDE prescription.  They devise the following formula which matches their results:

\begin{equation*}
\begin{aligned}
\dot N  = 1.1 F f_b \ln \left(0.22 \frac{M_{BH}}{M_*}\right ) \left(\frac{M_{BH}}{10^3 M_\odot}\right ) \left(\frac{\rho}{10^7 M_\odot {\rm pc}^{-3}} \right) \\ \times   \left(\frac{100 {\rm km s}^{-1}}{\sigma} \right )^3   {\rm Myr} ^{-1}
\end{aligned}
\end{equation*}

where $F$ is a numerical prefactor set to 0.8 and $f_b$ is the fraction of bound stars, which is assumed to be 0.2.  This analytic rate matches their simulations well, and while the simulations were only run for black hole masses growth up to  $\sim 3000 M_\odot$ for computational reasons, they suggest the mass growth will scale linearly with the TDE rate, which will remain as calculated above as long as the physical conditions  (e.g. density, velocity dispersion) of the cluster are unchanging.  These simulations assume old stellar ages and therefore do not include effects from supernovae or other stellar feedback in the cluster.  They also do not include the effect of multiple black holes, which could increase (by increasing the cross-section) or decrease (by scattering objects out of the cluster) the TDE rate.  Overall, both rate estimates come from different rationales and are in rough agreement with each other, so they can be used as a guideline.

\begin{figure}
\includegraphics[width=0.45\textwidth]{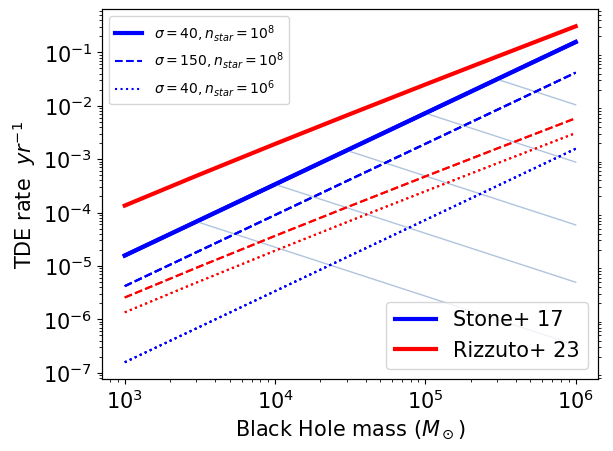}
\includegraphics[width=0.45\textwidth]{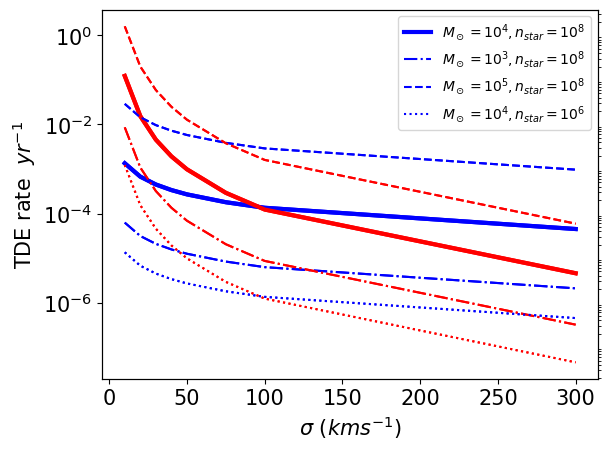}

\caption{ Rates of TDEs for models by \citet{Stone17} (blue) and \citet{Rizzuto23} (red).   Fiducial models shown in thick lines. {\em Top:} TDE rate vs black hole mass, with fixed velocity dispersion of $\sigma = 40$ km s$^{-1}$ (thick solid), $\sigma = 150$ km s$^{-1}$ (dashed), and with stellar density $n_{star} = 10^6 M_\odot$ pc$^{-3}$ (dotted).  Light  grey lines extending from the fiducial blue model indicate the rate in the instance of a transition from a full to empty loss cone, for various values of the critical mass $M_c$.  {\em Bottom:  } TDE rate vs stellar velocity dispersion, with fixed black hole mass of $10^4$ M$_\odot$ (thick solid), $10^5$ M$_\odot$ (dashed), $10^3$ M$_\odot$ (dot-dashed), and stellar density $n_{star} = 10^6 M_\odot$ pc$^{-3}$ (dotted).
\label{fig:tderates}
}
\end{figure}

Using these two rate estimates, I compute expected TDE rates in LRD environments.  The stellar densities in LRDs are estimated to be in the realm of $10^6 - 10^8 M_\odot$ pc$^{-3}$, and I use $10^8$ as a fiducial value \citep{Guia24}.  Using a canonical velocity dispersion for cluster collapse estimated at 40 km s$^{-1}$ \citep{Miller12}    and a mean stellar mass of 1 M$_\odot$, I plot the TDE rate in Figure \ref{fig:tderates} as a function of black hole mass (top) and stellar velocity dispersion (bottom)  (thick lines).   In the top plot the velocity dispersion is fixed at 40 km s$^{-1}$, while the bottom plot has a fixed M$_{\rm BH} = 10^4$ M$_\odot$.  The rate estimate by \citet{Stone17} is plotted in  blue and \citet{Rizzuto23} in  red.  A rate of  $10^{-4}$ per year is very reasonable within the range of $10^3 M_\odot < M_{\rm BH} < 10^5 M_\odot$ and/or velocity dispersions of 100 - 150 km s$^{-1}$.  There are no direct constraints on stellar velocity dispersions in LRDs, but it is plausible that with such high densities the values would be large. 

 Shown in the dashed, dotted, and dot-dashed lines in Figure \ref{fig:tderates} are rates calculated bracketing a reasonable LRD parameter space.  The dotted lines display a fiducial stellar density of $10^6 M_\odot$ pc$^{-3}$ rather than $10^8$, and show that for the lower bound of estimated densities the TDE rate may still reach $10^{-4}$ year$^{-1}$ but only for larger black holes and lower velocity dispersions.  The dashed and dot-dashed lines show extrema in velocity dispersion (150 km s$^{-1}$, upper panel) and black hole mass ($10^3$ and $10^5$M$_\odot$, respectively (lower panel).

 As mentioned above, the tidal capture prediction of \citet{Stone17} is only valid for the full loss-cone assumption.  At some critical mass $M_c$ the loss cone transitions from full to empty, at which point the rate of TDEs will scale as M$_{\rm BH}^{-11/12}$.  We include this calculation in the form of light grey downward-sloping lines in Figure \ref{fig:tderates} for a range of $M_c$ values.  (We include them on the fiducial calculation only so as not to further distract from the results of the figure.)   In many cases the  TDE rate increases with increasing black hole mass and decreasing velocity dispersion, and a rate of $10^{-4}$ year$^{-1}$  is unlikely to be reached in a case of small black hole mass and/or high velocity dispersion.

\section{Conclusions and Caveats}\label{sec:concl}

The plethora of LRDs discovered by JWST has challenged models of black hole and galaxy evolution.  In this paper I suggest that a primary origin for the emission from LRDs is due to tidal disruption events occurring during the process of forming the seeds of supermassive black holes.  The predicted number density of dense stellar clusters undergoing a runaway collapse is about $10^4$ times higher than that of LRDs, implying a TDE rate of $10^{-4}$ per year.   Other common attributes include a spectral rise in the UV/optical, broad H$\alpha$, and X-ray weakness.  The broad H$\alpha$ emission from TDEs is not correlated with black hole mass (unlike in AGN), providing an explanation for the proposed overmassive nature of these high redshift black holes relative to their host.  The H$\alpha$ lines also show absorption features \citep{Matthee24}, which could be due to tidal streams of disrupted stars. X-ray weakness is thought to be a sign of  super-Eddington accretion, likely combined with a viewing angle effect, and is thought to occur in TDEs.   In addition, when TDEs do show X-rays they are soft, which may be redshifted out of the Chandra detection range.

This model fails to directly address the red colors of LRDs.  The red part of the SED is likely at least somewhat dominated by stellar light, as evidenced by a common feature of a Balmer break \citep{Setton24} (although see \citet{Inayoshi24}).  LRDs do not characteristically show evidence of warm or hot dust (though see \citet{Barro24}), and therefore a standard AGN model (consisting of a torus) cannot apply.   Another common thorn in the side of LRD scientists is the number densities of these objects.  If LRDs are assumed to be dominated by AGN, their number densities are not far from those of all galaxies in the universe, implying a near 100\% AGN occupation fraction at $z \sim 7$ for the brightest galaxies ($L_{\rm bol} > 10^{45}$ erg s$^{-1}$).  Conversely, if most or all of the light is dominated by star formation, LRDs on their own reach or exceed the predicted stellar mass functions for all galaxies at high redshift in $\Lambda$CDM \citep{MBK23,Akins24}. 

These issues can  be avoided if one assumes a non-standard initial mass function (IMF).   A top-heavy IMF would result in higher luminosity stars, and an overall decrease in stellar mass estimates.   The IMF has been shown to vary with density and metallicity \citep{Marks12}, which are both extreme in the case of LRDs.   \citet{Jerabkova17} has shown that star clusters with top-heavy IMFs can reach up to quasar luminosities (i.e. $10^{46}$ erg s$^{-1}$ for a stellar mass of $10^9 M_\odot$), with a mass to light ratio of $10^{-3}$, for the first 10 Myr of their lifetimes. Therefore,  I  propose that the source of the light is stellar, originating in the dense nuclear cluster surrounding the IMBH, and with a top-heavy IMF.   The whole issue of too bright / too many stars is solved if there are actually fewer stars of higher masses.  A top-heavy IMF from a metal-poor population could sufficiently produce the red stellar light, explain a lack of dust (if the environment is very metal-poor), and solve the number density problem.  More massive stars would also result in more luminous TDEs.  SED models of LRDs with varying IMFs are needed to confirm this suggestion.

LRDs are not all identical, and there is no need for one overarching theory to explain their existence.  For example, some objects show strong AGN signatures \citep[see][]{Labbe24,Wang24} which likely dominate the SEDs.   Therefore, TDEs occurring in runaway collapsing clusters may explain a subset of LRDs, and canonical AGN  and other phenomena the others.  One way to verify the TDE hypothesis is to search for strong variability in the rest-frame UV (caused by TDE emission) but less in the rest-frame optical/NIR (dominated by stellar light).  TDEs have well-known decays in their light curves which can be confirmed fairly straightforwardly.   TDEs have also been shown to exhibit decaying luminosity in coronal lines on timescales of  $\sim$years \citep{Wang12}, which could be observed with spectroscopic follow-up.   The caveat is that due to time dilation timescales are lengthened by a factor of $(1+z)$, so decays will occur approximately 6-8 times more slowly than seen in the local universe.   Initial searches for variability of LRDs have found none as of yet \citep{Kokubo24,Tee24}, highlighting that these objects may have a diverse taxonomy.   Follow-up with JWST at appropriate timescales is crucial for verifying the TDE hypothesis.

\begin{acknowledgments}

 JMB is grateful for the support of NSF awards AST-2107764 and AST-2219090.   The conference Massive Black Holes in the First Billion Years in Kinsale, Ireland, was instrumental in inspiring this paper, especially conversations with Silvia Bonoli,  Suvi Gezari and Rosa Valiante.   JMB also thanks Jenny Greene, Ben Keller, Dale Kocevski, Erini Lambrides,  Mallory Molina, and the anonymous referee for helpful insights.  This research was conducted on the stolen lands of the Blackfoot, Umatilla, Crow, Cheyenne, and Salish peoples.
 
 \end{acknowledgments}

\bibliographystyle{aasjournal}

\end{document}